# Local negative magnetic permeability and possibility of observation of breather excitations in magnetic metamaterials

## M.M. Bogdan, O.V. Charkina

*B. Verkin Institute for Low Temperature Physics and Engineering of the National Academy of Sciences of Ukraine, 47 Nauky Ave., Kharkiv, 61103, Ukraine*

E-mail*:* m_m_bogdan@ukr.net , charkina@ilt.kharkov.ua

It is shown that the long-wave dynamics and magnetic properties of one-dimensional systems constructed of the inductively and capacitively coupled split-ring resonators are described by the regularized nonlinear dispersive Klein-Gordon equations. It is found that in such systems a highfrequency magnetic field excites dynamic solitons on a "pedestal" – stable breathers, oscillating in anti-phase with respect to the background of uniform oscillations, which means the existence of regions with a negative magnetic permeability in the system. If supplemented by a medium with negative permittivity, such a system forms a "left-handed" metamaterial in which the regions with the breather excitations are transparent to electromagnetic radiation. This makes it possible to observe them experimentally.

## 1. Introduction

Natural and artificial materials with negative permeability are receiving ever-increasing attention in the physics of magnetism over the last decade [1, 2]. This interest is stirred by the emergence and rapid development of a new direction associated with the creation of metamaterials – artificial media with negative refractive index for electromagnetic waves in the gigahertz and terahertz ranges [3, 4]. In such metamaterials, the permittivity and permeability can simultaneously become negative within the specified frequency range. This makes possible the propagation of waves with unusual "left-handed" properties as predicted by Veselago [5]. The distance between the structural units forming the metamaterial is much smaller than the wavelength of the electromagnetic wave, so it can be considered as a continuous medium. The conductive media with negative dielectric constant [6, 7] and natural compounds with negative magnetic permeability have been studied separately in detail [1]. However, only relatively recently, the theoretical ideas by Veselago have been realized experimentally by creating an artificial medium composed of two subsystems: A lattice of metal rods [6] and a magnetic metamaterial formed by chains of open-ring resonators inductively coupled to each other [8].

Depending on the dimensions of their structural units – ring resonators, the magnetic metamaterials have different threshold frequencies starting from which they exhibit negative magnetic permeability. At frequencies below the threshold, the magnetic response of such systems to uniform pumping is positive [8, 9].

New properties of the magnetic metamaterials arise when nonlinear elements are incorporated into their structural units [10, 11]. In this case, an essentially uniform alternating magnetic field $H(\tau) = H_0 \cos\Omega\tau$ induces the emf $E(\tau) = E_0 \sin\Omega\tau$ in the magnetic metamaterial, the appearance of which can lead to the excitation of strongly nonlinear oscillations. The amplitude of this emf is equal to $E_0 = \mu_0 \Omega S H_0$, where $\mu_0 = 4\pi \cdot 10^{-7}$ H/m is the magnetic permeability of vacuum and $S$ is the area of an open-ring resonator [9]. Lazarides and co-workers have shown theoretically [12, 13] that in 1D case the electrical oscillations of inductively-coupled open-ring resonators with nonlinear elements are described by the following discrete equations:

$$\begin{cases} L\dfrac{dI_n}{d\tau} + RI_n + U_n - M\left(\dfrac{dI_{n-1}}{d\tau} + \dfrac{dI_{n+1}}{d\tau}\right) = E(\tau) \\ \dfrac{dQ_n}{d\tau} = I_n \end{cases}, \quad (1)$$

where $I_n$ is the ac current in the circuit, $Q_n$ is the charge, $L$ and $M$ are the self-inductance and mutual inductance, $R$ is the active resistance, and the nonlinear dependence of the capacitive voltage on the charge $U_n = f(Q_n)$ is provided, for instance, by diodes in the circuit or by a dielectric with the Kerr nonlinearity in the gap of the ring. Numerically, it was shown that discrete equations (1) have solutions which are highly localized in space and periodic in time and correspond to discrete breathers [14]. In particular, it was shown that in a small region of the order of several lattice constants, a discrete breather with amplitude of opposite sign with respect to the applied field can be excited [13]. This implies that in the area encompassing several resonators, the metamaterial exhibits diamagnetic response. Unfortunately, for such a strong localization it is difficult to discuss the negative permeability of the metamaterial since this is a macroscopic characteristic of a continuous medium. Recently, attempts have been made to find analytical expressions for the discrete nonlinear excitations in magnetic metamaterials. However, microscopic theoretical description of the dynamics of nonlinear metamaterials in the highly discrete case appears to be a challenging task for which numerical methods are mainly applied [12, 13, 15].

This paper proposes an analytical approach to the description of the dynamic characteristics of breather oscillations in metamaterials with both inductive and capacitive coupling between the resonators taken into account. It is theoretically shown that in such systems, the finite regions with negative magnetic permeability can exist, which are related to the presence of magnetically-induced breathers in these regions of the metamaterial.

## 2. Forced breather oscillations in a magnetic metamaterial

The metamaterial modification proposed in this work includes an additional effective capacitive coupling $C$ between the resonators as compared to model (1). Such a coupling is almost always present in metamaterials [16].

Let us write, similar to Eq. (1), the system of equations for the voltage in the circuit

$$\begin{cases} \tilde{L}\dfrac{dI_n}{d\tau} + RI_n + f(Q_n) - \dfrac{2}{C}Q_n - \dfrac{1}{C}(Q_{n-1} + Q_{n+1} - 2Q_n) - M\left(\dfrac{dI_{n-1}}{d\tau} + \dfrac{dI_{n+1}}{d\tau} - 2\dfrac{dI_n}{d\tau}\right) = E(\tau) \\ I_n = \dfrac{dQ_n}{d\tau} \end{cases} \quad (2)$$

where the second differences are grouped together in the expressions for capacitive and inductive coupling. This results in the renormalization of the coefficients in the inductive terms related to node $n$: $\tilde{L} = L - 2M$. We consider a system of planar open rings, for which the parameter $M > 0$. The influence of non-linear elements in the metamaterial is described, as a rule, by the first terms in the expansion of the dependence $f(Q_n) = Q_n(1 - \eta(Q_n/Q_c)^2)/C_0$, where $C_0$ is the linear capacitance of an open ring considered as a tank circuit, $Q_c$ is the characteristic charge, and $\eta$ is a numerical coefficient of cubic nonlinearity. Then the system of Eq. (2) can be written as a single equation for the charge variable $Q_n$

$$\tilde{L}\frac{d^2 Q_n}{d\tau^2} + R\frac{dQ_n}{d\tau} + \frac{1}{\tilde{C}}Q_n - \frac{\sigma}{C_0}Q_n^3 - \frac{1}{C}(Q_{n+1} - 2Q_n + Q_{n-1}) - $$
$$- M\frac{d^2}{d\tau^2}(Q_{n+1} - 2Q_n + Q_{n-1}) = E(\tau) \tag{3}$$

where $\tilde{C}^{-1} = C_0^{-1} - 2C^{-1}$.

Let us consider the long-wave excitations propagating in the metamaterial considered as a continuous medium. In the long-wavelength limit, when the distance between the open rings $d$ is much smaller than the oscillation wavelength, the second differences are converted into the second derivatives with respect to the coordinates, and the second differences of the second derivative of charge with respect to the time are readily seen to be a direct implementation of the fourth mixed spatio-temporal derivative. Then the equation for the dimensionless charge $u_n = Q_n/Q_c$ is reduced to the nonlinear Klein-Gordon equation with the mixed fourth derivative, the so-called regularized Klein-Gordon equation (RKGE) [17], with additional terms describing the damping (active resistance) and external variable force (emf)

$$u_{tt} + \lambda u_t - u_{xx} - \beta u_{xxtt} + u - \eta u^3 = e_0 \sin(\omega t). \tag{4}$$

The variable $u(x,t)$ depends on the dimensionless coordinate $x = nd/l_c$ and the time $t = \omega_0 \tau$, where $\omega_0 = (\tilde{L}\tilde{C})^{-1/2}$ is the natural frequency of the ring circuit, $l_c = \omega_c d/\omega_0 = d\sqrt{\tilde{C}/C} \gg d$ is the characteristic length, $\lambda = R\sqrt{\tilde{C}/\tilde{L}}$ is the oscillation damping coefficient, $\omega = \Omega/\omega_0$ and $e_0 = E_0 \tilde{C}/Q_c$ are the dimensionless frequency and pumping amplitude. After substituting $E_0$ into the expression for the last parameter, we obtain an expression, which provides an explicit dependence on the magnetic field and the pumping frequency

$$e_0 = \frac{\mu_0 H_0 \Omega S \tilde{C}}{Q_c} = \frac{\mu_0 S}{Q_c}\sqrt{\frac{\tilde{C}}{\tilde{L}}}\omega H_0. \tag{5}$$

As can be seen from Eq. (4), the inductive coupling, which is described by the term with the mixed fourth derivative, inevitably brings higher-dispersion effects into the system. The dispersion parameter is determined by the product of the ratios of the inductance and capacitance coefficients

$$\beta = \frac{MC}{\tilde{L}\tilde{C}} = \frac{M}{\tilde{L}}\left(\frac{d}{l_c}\right)^2 \tag{6}$$

and, within the long-wavelength approximation, it is natural to assume it to be below unity. If damping and external pumping are neglected, Eq. (4) becomes literally the regularized nonlinear Klein-Gordon equation with cubic nonlinearity [17]

$$u_{tt} - u_{xx} - \beta u_{xxtt} + u - \eta u^3 = 0. \tag{7}$$

The problem of small-amplitude self-localized oscillations (breathers) for the regularized sine-Gordon equation has been solved in Ref. [17], and its results can be directly applied (up to the factor of 1/6 in the expansion of sinusoidal force with respect to the general coefficient $\eta$) to Eq. (7). In particular, for a small-amplitude RKGE breather, all the conclusions of Ref. [17] regarding the behavior of the fundamental harmonic of the breather remain valid, including the specific feature of its spatial localization – the disappearance of the amplitude dependence of the effective width of the breather in the limit of the dispersion parameter $\beta \to 1$. This property remains essential also for the study of forced oscillations.

Returning to the original problem (4) of the nonlinear dynamics of magnetic metamaterials under external pumping and dissipation, let us investigate the regimes of stable localized forced oscillations of the breather type and find the nonlinear response of the magnetic metamaterial with an excited breather. In what follows we neglect the losses in the active resistance. Let us consider the external pumping frequency below $\omega_0$ – the minimum frequency limit for the linear oscillations of the system. Then, as in Ref. 17, by using the asymptotic Kosevich-Kovalev procedure, i.e., by plugging the solution represented as an expansion with respect to odd temporal harmonics with the pumping frequency $\omega$ and assuming an hierarchy of smallness with respect to the parameter of frequency separation from the spectral edge $\kappa = \sqrt{1-\omega^2} \ll 1$

$$u(x,t) = A(x)\{[1+\kappa^2 B(x)]\sin\omega t + \kappa^2 C(x)\sin 3\omega t + ...\}, \quad (8)$$

we obtain a nonlinear equation for the amplitude of the fundamental harmonic

$$(1-\beta\omega^2)A_{xx} - (1-\omega^2)A + \frac{3}{4}\eta A^3 + e_0 = 0. \quad (9)$$

For all the terms to be of the same order of smallness, it is necessary that the amplitude of the function $A(x)$ is of the order of $\kappa$ and the region of localization is of the order of $1/\kappa$, while the pumping amplitude $e_0$ is of the order of $\kappa^{3/2}$, which can be always achieved by varying the magnetic field $H_0$.

Equation (9) has three spatially homogeneous solutions which are the roots of the cubic equation

$$(1-\omega^2)a - \frac{3}{4}\eta a^3 = e_0 \quad (10)$$

It is known that for small $e_0$ all three solutions of the cubic equation are real, and, moreover, two of them are stable. In Ref. 18 it was shown that the formation of a soliton is possible on the background of a stable solution with the lowest possible amplitude $a$. The explicit dependence of the amplitude of the homogeneous oscillation $a$ on the pumping parameters $\omega$ and $e_0$ is given by

$$a = \frac{2}{3\sqrt{\eta}}\sqrt{1-\omega^2}\left(\cos\frac{\gamma}{3} - \sqrt{3}\sin\frac{\gamma}{3}\right), \quad \gamma = \arccos\left(\frac{9}{4}\frac{\sqrt{\eta}e_0}{(1-\omega^2)^{3/2}}\right) \quad (11)$$

Let us separate the homogeneous part $a$ of the solution for the amplitude $A(x) = a + A_s(x)$. After substituting it in Eq. (9), we obtain the equation for the pure soliton dependence $A_s(x)$, which satisfies the boundary conditions vanishing at infinity. This equation can be integrated explicitly and the breather solutions on a homogeneous background – solitons on a "pedestal" – can be finally obtained

$$A(x) = A_{\pm}(x) = a\left[1 + \frac{2\sinh^2(\alpha)}{1 \pm \cosh(\alpha)\cosh(x/l)}\right]. \quad (12)$$

Here we introduced the notation

$$2\sinh^2(\alpha) = \frac{4}{3\eta a^2 l^2}, \qquad l^2 = \frac{1 - \beta\omega^2}{1 - \omega^2 - \frac{9}{4}\eta a^2}, \quad (13)$$

where the parameter $l$ is the characteristic size of the localization of the breather. Note that the solution exists for the frequencies which are very close, but strictly below the resonance $\omega \leq \omega_c < 1$; at the frequency $\omega_c$ the parameter $l$ tends to infinity.

As can be seen from expression (12), the amplitude in the center of the breather can be either positive or negative. Stability of such solutions within the framework of the nonlinear Schrodinger equation with external pumping has been studied in Ref. [18]. It has been shown that a completely positive solution with smaller amplitude is unstable, while the solution $A_-(x)$ with negative amplitude at the center is stable under small perturbations. Further on, an important feature of the function $A_-(x)$ for analysis and applications is the fact that it depends not only on the coordinate $x$, but also on the frequency $\omega$ as a parameter for a fixed value of $e_0$. We will emphasize this by the notation $A_-(x) \equiv A_-(x,\omega) \equiv a(\omega) \cdot f(x,\omega)$.

### 3. Local negative magnetic permeability and observation of breathers in metamaterials

Having the explicit solution, we can analytically find the permeability of the metamaterial containing the magnetic breather – a dynamic soliton on a "pedestal." Following the analysis of the general equations for magnetic induction [12], it is straightforward to conclude that a localized region of negative magnetic permeability should exist in the metamaterial. The expression for magnetic induction has the standard form

$$B = \mu_0(H(\tau) + M(x,\tau)), \quad (14)$$

where the magnetization of the metamaterial $M(x,\tau)$ at the point $x$ is determined through the strength of the current $I(x,\tau)$ in the ring and is equal $M(x,\tau) = Sd^{-3}\partial Q(x,\tau)/\partial\tau$ [9]. After plugging the solution for $Q(x,\tau) = Q_c A_-(x)\sin\Omega\tau$, we obtain the expression for magnetic induction

$$B = \mu_0 \mu_r(x,\Omega) H_0 \cos\Omega\tau, \quad (15)$$

with the relative magnetic permeability and susceptibility

$$\mu_r(x,\Omega) = 1 + \chi(x,\Omega), \quad (16)$$

$$\chi(x,\Omega) = \rho A_-(x,\Omega) = \rho a f(x,\Omega), \quad (17)$$

where the dimensionless parameter $\rho$ is equal

$$\rho \equiv \frac{S\Omega}{H_0 d^3} Q_c. \quad (18)$$

It can be seen from Eqs. (16) and (17) that the possibility of appearance of a region with negative values of permeability $\mu_r(x,\Omega)$ in a metamaterial with an excited breather significantly depends not only on the value of $\rho$, but also on the contribution of the fundamental harmonic

6amplitude since the susceptibility $\chi(x,\Omega)$ is of the order of $\rho a$. To estimate the amplitude of the homogeneous solution $a$, in the first approximation we can take $a = e_0/\kappa^2$, and, after plugging $e_0$ from Eq. (5) into it and using the definition of $\omega_0^2$, we finally obtain

$$a \approx \frac{\mu_0 H_0 \Omega S}{Q_c \tilde{L}(\omega_0^2 - \Omega^2)}. \tag{19}$$

Substituting Eqs. (18) and (19) into Eq. (17), we obtain an estimate for the susceptibility

$$\chi(x,\Omega) \approx \frac{\Omega^2}{\omega_0^2 - \Omega^2} \frac{\mu_0 S^2}{\tilde{L} d^3} f(x,\Omega). \tag{20}$$

For a ring resonator, the area parameter can be estimated by a simple expression $S \approx \pi d^2$, so we finally obtain

$$\chi(x,\omega) = \rho A_-(x,\omega) \approx \frac{\pi^2 \omega^2}{1-\omega^2} \frac{\mu_0 d}{\tilde{L}} f(x,\omega). \tag{21}$$

Nanomagnetic metamaterials for the terahertz range ($\omega_0 = 10^{12}$ Hz) have the typical dimensions of structural elements and the distance d between them about $d \approx 5 \cdot 10^{-6}$ m on average [4] and the inductance $\tilde{L} \approx 6 \cdot 10^{-12}$ H [12]. This implies that the quantity $\mu_0 d/\tilde{L} \approx 1$, and the product of the dimensionless parameters $\rho a \gg 1$ due to the proximity of $\omega^2$ to unity. Thus, the sign of the magnetic permeability is completely determined by the dependence of $A_-(x,\omega)$, and the diamagnetic response in the region of localization of the breather substantially exceeds the positive response of the homogeneous background. This dependence on the dimensionless coordinate along the metamaterial and the pumping frequency normalized to $\omega_0$ is shown in Fig. 1(a) for a fixed value of the magnetic field $H_0$. Besides the function $a(\omega)$, also the parameters $l(\omega)$ and $\alpha(\omega)$ in Eq. (10), solution of Eq. (9), contribute to the frequency dependence. Fig. 1(b) shows the dependence of $A_-(x,\omega_*)$ on the coordinate for a fixed frequency ($\omega_* = 0.95$). It can be seen that region I where the function $A_-(x,\omega)$ is negative, i.e., actually the region of negative permeability, lies in the interval $[-x_0, x_0]$, the size of which depends on frequency through the parameters of the solution $x_0 = l \cdot \mathrm{arcosh}(\cosh 2\alpha/\cosh \alpha)$. Outside this range (region II) the permeability is positive. Note that in the limit $\beta \to 1$, the region of negative permeability is almost independent of the small parameter $\kappa$. This property may seem important, because it means that when the region of continuous spectrum shrinks, the region of negative magnetic permeability remains macroscopic and finite.

Using numerical simulations, we investigated the dynamic stability of the forced breather modes – solitons on a "pedestal" – within the initial RKGE with external pumping, Eq. (4). It turned out that there is a range of parameter values, which are acceptable within the above-developed theory, where the regimes of forced oscillations realizing the local negative magnetic permeability are stable. Fig. 2 shows such a stable breather mode oscillating in strict anti-phase with respect to the uniform background for the values of the dispersion parameter $\beta = 1/8$, the nonlinearity coefficient $\eta = 1/6$, and the pumping frequency $\omega = 0.95$ and amplitude $e_0 = 0.01$. On the other hand, it turned out that the solution loses its stability in the frequency limit where $l(\omega) \to \infty$ and thus has a limited region of stability. Nevertheless, this occurs at frequencies very close to $\omega_0$.

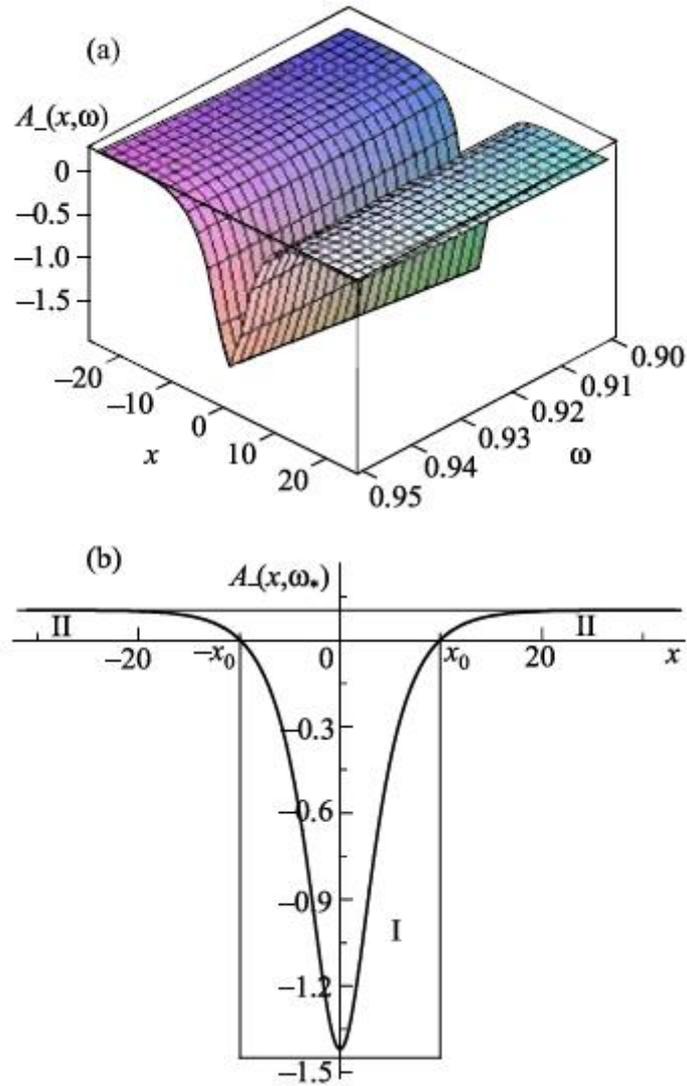

**FIG. 1.** *Nonlinear response of the metamaterial containing a dynamic soliton on a "pedestal": The dependence of the soliton amplitude on the coordinate and the pumping frequency at a fixed value of magnetic field (a); the soliton profile at the fixed frequency $\omega_* = 0.95$, I is the region of negative permeability, II is the region of positive permeability (b).*

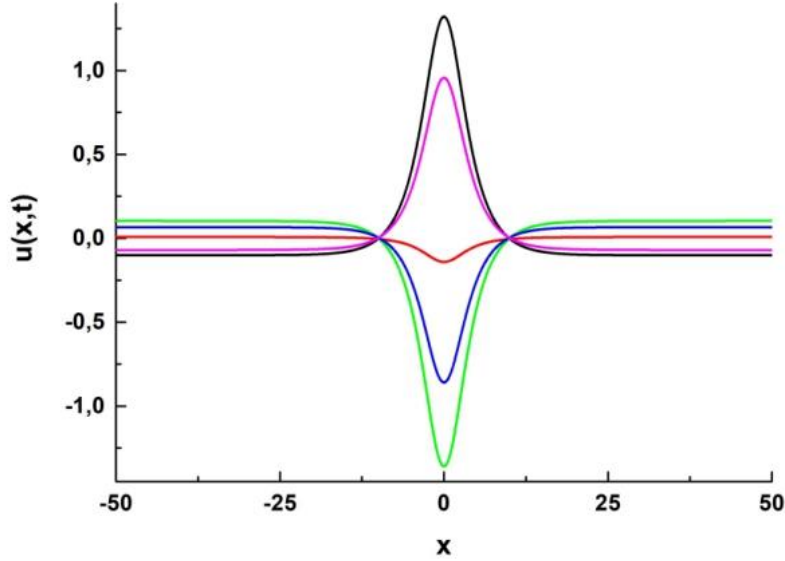

**FIG. 2.** *Stable regime of forced breather oscillations for the dispersion parameter* $\beta = 1/8,$ *and the pumping frequency* $\omega = 0.95,$ *and amplitude* $e_0 = 0.01$.

Thus, the excitation of breather oscillations in metamaterials can lead to the existence of a nonlinear local diamagnetic response of the metamaterial to an external high-frequency field. This effect makes it possible to observe the magnetic breather excitations. If the above-considered nonlinear chain of coupled open-ring resonators is complemented by a lattice of conductive rods which exhibits negative permittivity $\varepsilon$ in a given frequency range, then the refractive index $n = \sqrt{\varepsilon\mu}$ is negative in the region of localization of the magnetic breather and purely imaginary everywhere outside this area. The corresponding region of the metamaterial becomes transparent to the passage of an electromagnetic wave with the magnetic field vector component perpendicular to the plane of the open ring resonators. Two dimensional metamaterial exhibiting similar properties can be constructed from one-dimensional non-interacting chains of planar-oriented ring resonators. The localization of breathers in the chains is independent, so the pattern of intensity distribution for a wave transmitted through the two dimensional metamaterial can be significantly heterogeneous and exhibit peaks at the locations of breathers. Such distribution will strongly depend on the pumping frequency and amplitude. Note that similar intriguing phenomena have been observed in a two-dimensional nonlinear metamaterial [19] and, in our opinion, can be explained by the above-described mechanism associated with the excitation of breathers – dynamic solitons on a "pedestal" – in such a metamaterial.

In conclusion, it should be noted that the equation describing purely magnetoinductive excitations (without considering the capacitive coupling) in the long-wavelength limit has the form

$$u_{tt} - \beta u_{xxtt} + u - \eta u^3 = 0 \qquad (22)$$

The breather eigenmodes in such a regularized equation without high-frequency pumping and dissipation have been studied in detail previously [20] outside the context of the theory of metamaterials. In the discrete model of a metamaterial under external pumping (1), the existence of strongly localized dynamic solitons on a "pedestal" has been found in Ref. 13. Analytical description of these forced long-wavelength breather oscillations and the possibility of their excitation with pumping and dissipation taken into account will be discussed in a separate paper.